\documentclass[runningheads]{llncs}
\usepackage{graphicx}
\usepackage{algorithm}
\usepackage[noend]{algcompatible}
\newcommand{\algorithmicinput}{\textbf{input}}
\newcommand{\INPUT}{\item[\algorithmicinput]}
\usepackage{wrapfig}
\usepackage{amsmath}

\begin{document}
\title{BayesNetCNN: incorporating uncertainty in neural networks for image-based classification tasks}
\titlerunning{Bayesian NN for Alzheimer prediction on JD images}
%
\author{Matteo Ferrante\inst{1} \and
Tommaso Boccato\inst{1} \and
Nicola Toschi\inst{1,2}}
\authorrunning{Ferrante et al.}
%
\institute{Department of Biomedicine and Prevention, University of Rome Tor Vergata (IT) \and Martinos Center For Biomedical Imaging, MGH and Harvard Medical School (USA) \and 
\\
\email{\{matteo.ferrante,tommaso.boccato\, nicola.toschi}@uniroma2.it}
\maketitle              
\begin{abstract}
The willingness to trust predictions formulated by automatic algorithms is key in a vast number of domains. However, a vast number of deep architectures are only able to formulate predictions without an associated uncertainty. In this paper, we propose a method to convert a standard neural network into a Bayesian neural network and estimate the variability of predictions by sampling different networks similar to the original one at each forward pass. We couple our methods with a tunable rejection-based approach that employs only the fraction of the dataset that the model is able to classify with an uncertainty below a user-set threshold. We test our model in a large cohort of brain images from Alzheimer’s Disease patients, where we tackle discrimination of patients from healthy controls based on morphometric images only. We demonstrate how combining the estimated uncertainty with a rejection-based approach increases classification accuracy from $0.86$ to $0.95$ while retaining $75\%$ of the test set. In addition, the model can select cases to be recommended for manual evaluation based on excessive uncertainty. We believe that being able to estimate the uncertainty of a prediction, along with tools that can modulate the behavior of the network to a degree of confidence that the user is informed about (and comfortable with) can represent a crucial step in the direction of user compliance and easier integration of deep learning tools into everyday tasks currently performed by human operators.

\end{abstract}

\section{Introduction}

The willingness to trust predictions formulated by automatic algorithms is key in a vast number of domains. In addition to questions of ethics and responsibility, it is important to note that, whilst extremely powerful, a vast number of deep architectures are only able to formulate predictions without an associated uncertainty. This shortcoming critically reduces user compliance even when explainability techniques are used, and this issue is particularly sensitive when deep learning techniques are employed e.g. in the medical diagnosis field.
Being able to produce a measure of system confidence in its prediction can really improve the trustability in the deep learning tool as a recommendation machine able to improve the workflow of physicians. 
Alzheimer's  disease (AD) is one of the most critical public health concerns of our time. Due to life expectancy increases and better professional care more and more people reach older ages but are often affected by degenerative brain disorders like AD, which is a severe form of dementia \cite{alzheimer}.
Principal symptoms are progressive memory loss, difficulties in normal-life activities, language disorders, disorientation, and, in general, a decrease in cognitive functions. One of the most important risk factors is age, while in some cases specific genetic mutations are responsible for pathology onset, which however can also be related to comorbidities. As a progressive degenerative pathology, AD is usually preceded by a different condition called mild cognitive impairment (MCI), with less intense symptoms that often, but not always, evolve into AD, which has no cure. Many theories about the etiopathogenesis of AD exist, several of which are linked to an alteration in the metabolism of the precursor protein of beta-amyloid. The latter’s metabolism slowly changes over the course of the years, leading to the formation of neurotoxic substances which slowly accumulate in the brain. The causal relationship between beta-myeloid metabolism and clinical AD presentation is the object of intense research \cite{simeon,Toschi,Hampel2018-kv,Hampel2020-rp}. In clinical practice, AD diagnosis is based on the symptoms and commonly confirmed using magnetic resonance imaging (MRI) or positron emission tomography (PET), which however leaves the clinician with a great deal of subjectivity and uncertainty to deal with when positioning a patient in the AD continuum.
For this reason, there is great interest in models able to detect and predict AD-related structural and functional changes. Deep learning models are able to usefully extract local and global features through convolutional layers and learn how to predict interesting outcomes, such as distinguishing healthy controls from AD patients or even MCI patients which will remain stable for those who will progress to AD \cite{alz_deep,simeon,cad,predict_alz}.
In this context, difficulties in accessing large-scale curated datasets and the need to work with multimodal high-dimensional data, call for particular attention to avoiding overfitting and increasing the reliability of automatic models, possibly including the output of uncertainty estimates which can be evaluated by neuroscientists and physicians. For those reasons, we propose a Hybrid Bayesian Neural Network in a framework where predicted probabilities are coupled with their uncertainties. To reduce the number of parameters, we propose a convolutional neural network based on depthwise separable convolutions. We trained our model on a subset of the Alzheimer's Disease Neuroimaging Initiative (ADNI) dataset using Jacobian Determinant images, that is, images where each voxel describes the change in the volume element resulting from nonlinear coregistration of the patient’s brain MRI into a standard space such as e.g. the Montreal Neurological Institute (MNI) T1-weighted template. This choice was made in order to isolate morphometric changes (such as e.g. cortical atrophy) from image intensity variations  \cite{morphometry}.
Once trained, we turned the last linear layer into a Bayesian neural network, replacing optimal values $w*$ with narrow parameter distribution $N(w^*,s)$. This means that instead of having a single weight value for each connection between two neurons from last layer to final output, a Gaussian distribution centered on the optimal value $w^*$ is used. Every time the network does an inference the weights of the last layer are sampled from those distributions. In this way is possible to obtain $N$ slightly different
networks which, in turn, allows to perform ensembling and hence provide an uncertainty estimate. The latter can also be thresholded in order to subset the data and increases prediction performance.

\section{Material and methods}

\begin{figure}[htbp]
    \centering
    \includegraphics[width=0.5\linewidth]{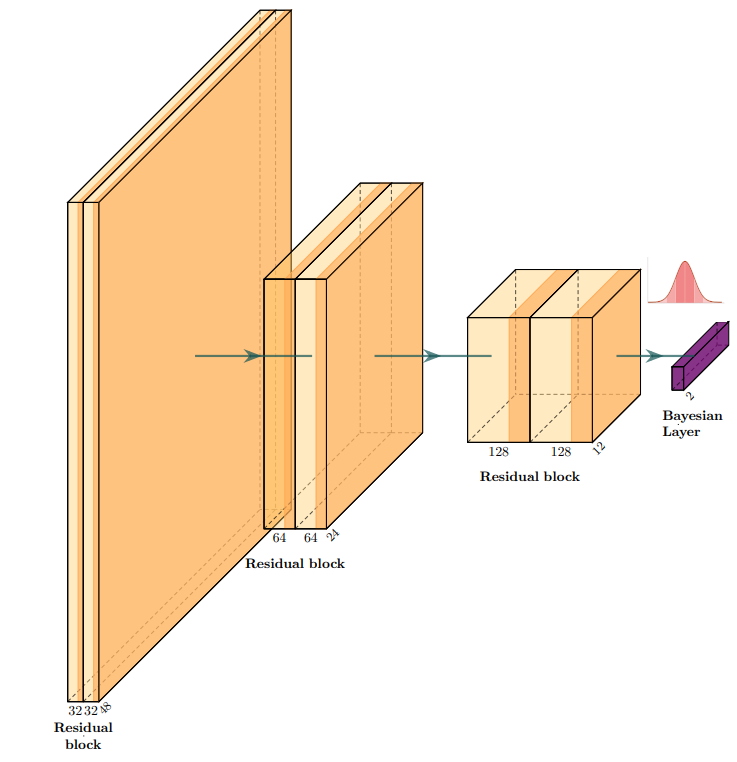}
    \caption{Architecture overview: Our model is a classifier based on residual convolutional blocks. Each block is composed of two depthwise separable convolutional layers to keep the number of parameters as low as possible to process 3D images. In correspondence with the gaussian distribution symbol we sample the network classifier weights $w\sim \mathcal{N}(w^*,s)$.}
    \label{fig:architecture}
\end{figure}

In this section, we describe the dataset and briefly revisit the theory behind Bayesian neural networks that justifies our approach.

\subsection{Dataset}
We selected a subset of 376 cases from the ADNI \cite{adni} dataset, composed of cases labeled as both healthy and AD and employed the Magnetization Prepared RApid Gradient Echo (MPRAGE), T1-weighted image only.T1-weighted (T1w) images were coregistered to the MNI template using linear initialization and a nonlinear warp, after which the Jacobian Determinant (JD) maps were computed by isolating the nonlinear part of the deformational field which takes the images from native space to standard space. We finally masked the deformation maps using the standard MNI brain mask. Registration procedures were performed using the ANTs package \cite{ants}.
 The high-dimensional nonlinear transformation (symmetric diffeomorphic normalization transformation) model was initialized through a generic linear transformation that consisted of the center of mass alignment, rigid, similarity, and fully affine transformations followed by nonlinear warps (metric: neighborhood cross-correlation, sampling: regular, gradient step size: 0.12, four multiresolution levels, smoothing sigmas: 3, 2, 1, 0 voxels in the reference image space, shrink factors: 6, 4, 2, 1 voxels. We also used histogram matching of images before registration and data winsorization with quantiles 0.001 and 0.999. The convergence criterion was set to be as follows: the slope of the normalized energy profile over the last 10 iterations < 10-8). Coregistration of all scans required approximately 19200 hours of CPU time on a high-performance parallel computing cluster.
Our final dataset consisted of 376 JD images, evenly distributed between AD and healthy cases. The dataset was split in an 80(train)/20(test) fashion, normalized globally, and cropped to a (96,96,96) size.

\begin{figure}[h!]
    \centering
    \includegraphics[width=0.9\linewidth]{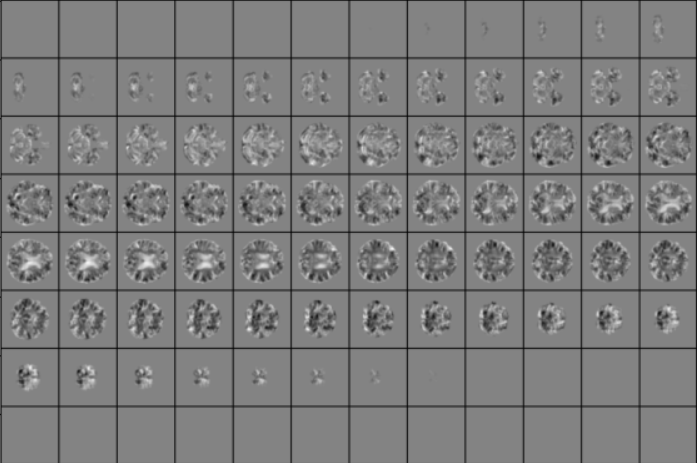}
    \caption{Example of slices for one random case in the test set. Local Jacobian Determinant images are normalized in the range [0,1].}
    \label{fig:architecture}
\end{figure}

\subsection{Bayesian Neural Network}
We briefly recap the theory behind Bayesian neural networks and then describe the architecture of our model and the training procedure.

The idea is that instead of estimating $w^*$ which minimizes the cost function, we
learn a weight distribution. This is equivalent to an infinite ensemble approach, which allows us to estimate the variance of the prediction, sampling a slightly different neural network each time we perform inference.
Instead of learning $w^*$ we learn the posterior $p(w|D)$, where $D$ represents the incoming data.
Our aim here is to perform inference as the average of different neural networks as described in Equation \ref{eq:1}.

\begin{equation} 
p(\hat{y}|D)=\int_w p(y|w)p(w|D)dw = E_{p(w|D)}(p(y|w))
\label{eq:1}
 \end{equation}

Where $p(\hat{y}|D)$ is probability of get the prediction $y$ given the data $D$, $p(y|w)$ the conditioned probability of get $y$ given network's weights $w$ and $p(w|D)$ the posterior probability of having weight $w$ given the data $D$.

To perform this computation we need the posterior $p(w|D)$ which can be rewritten using the Bayes theorem.
So $p(w|D)$ can be expressed by the likelihood $p(D|w)$ and the prior $p(w)$ but we also need the normalization term at the denominator which is computationally intractable.
At this point, several approaches exist to overcome this issue.
One popular approach is variational inference, trying to estimate $p(w|D)$ approximating this distribution with a parametrized distribution $q_\phi(z)$ that minimizes the Kullback-Leibler (KL) divergence with the target distribution.
Monte Carlo approaches are also possible, sampling points matching the required distribution as described in \cite{bayes_back,bayesian,variation_autoencodes,variational_dropout}.

This latter approach is anyway computationally intensive while using variational inference requires changes in the objective function that accomplish a new task described by a modification in the loss function. In this case, the standard loss function is augmented with the KL divergence between the distribution of the weights and the chosen prior which can make training unstable and longer.

Instead, we opted for a hybrid approach, first training a standard convolutional network and then turning the last layer weights into narrow Gaussian distributions centered with the optimal values $w^*$.
Assuming that optimal values can serve as the center of Gaussian Distribution and little deviation around these optimal values can represent similar networks this approach turns a standard neural network into a Bayesian one without the need for any added complexity during training time.

\begin{equation}
    p(w|D)= \frac{p(D|w)p(w)}{\int_{w'} p(D|w')p(w')dw'}
\end{equation}

\subsection{Neural network architecture}

Our base model is a residual convolutional neural network based on depthwise separable convolutions, which we implemented to reduce the number of parameters and the risk of overfitting. 3D depthwise separable convolutions are based on an hoc PyTorch implementation, using grouped convolutions with the group number set to the same value as the number of input channels, followed by a pointwise convolution with output channels. In other words, convolutions are first learned channelwise, and then information about the interaction between channels is taken into account by the second depthwise convolution for each point. This reduces the number of parameters from  $COK^3$ to $C(K^3+O)$. Here $C$ is the number of input channels, $K$ is the 3D kernel size, $O$ is the number of output channels. 
 Our model is composed of three residual blocks, and each block is composed of two depthwise separable convolutions with a PReLU activation function \cite{prelu}. Each block halves  the side dimension of the images. A flattening layer is followed by a linear layer for the first part of the training and then turned into a Bayesian linear layer replacing optimal values with Gaussian distributions. We used the Adam optimizer  \cite{adam} with a learning rate of $3e-4$ and trained our model for 5 epochs.
All implementation was built in python, using Pytorch, Monai \cite{monai} and Torchbnn libraries \cite{bayesian}.

After training the last layer weights were replaced by a set of Gaussians $w* \rightarrow N(w*,s)$ with s chosen to be small. We set $s=0.01$ in our experiments.
At each forward pass, the network processes information the standard way until reaching the last layer.
Here, a set of weights is sampled from $\title{w} \sim \mathcal{N}(w^*,s)$ generating a slightly different neural network.
Sampling $N$ networks in inference produce different estimation and let us the chance to estimate uncertainty about the output.

\subsection{Experiment}
Our model is trained to classify AD and healthy cases on JD images.
We ran inference on our test set with $N=100$, each time sampling the weights from their distributions.
We computed the \textit{softmax} of the output to obtain the probabilities and aggregate the results by means.
We also computed the standard deviation for each outcome probability.
Then, we set a set of thresholds on standard deviation to study the variation of performances.
Each estimation with a standard deviation is rejected. 
The idea is that we can set the threshold according to our needs. If we need higher accuracy and avoid uncertain estimation we can set a small value for the threshold $t$. In this case, we'll have fewer data accepted for estimation and more "rejected" cases to be reviewed manually. On the other side, we can retain most or all of the test dataset if we can accept more misclassified cases.

Algorithm \ref{alg:inference} describes the whole procedure.

\begin{algorithm}[H]
  \caption{Inference}
  \label{alg:inference}
  \begin{algorithmic}
  \INPUT{$x, N$ JD images, number of inference}
        \FOR{$i$ in range($N$)}
        \STATE Sample NN weights $w\sim \mathcal{N}(w^*,s)$
        \STATE Estimate output probabilities $p_i=f_w(x)$
        \ENDFOR
    \STATE Average prediction $p_{mean}=\frac{1}{N}\sum_{i} p_i$
    \STATE Compute $p_{std}$
    
  \end{algorithmic}
\end{algorithm}

\begin{algorithm}[H]
  \caption{Reject procedure }
  \label{alg:rejection}
  \begin{algorithmic}
  \INPUT{$p_{mean}, p_{std}$ for test set}
  \STATE keep=[]
        \FOR{sample in test set}
        \IF {$p_{std}<threshold$} 
            \STATE keep.append($p_{mean}$)
        \ELSE 
            \STATE Reject $p_{mean}$
        \ENDIF
        \ENDFOR
\STATE Evaluate keep

  \end{algorithmic}
\end{algorithm}

\subsection{Explainability}

In order to visualize the portion of the images which were weighed most by our model, we used the trulens library  \cite{trulens} implementation of integrated gradients in \cite{integrated_gradient}. 
A baseline $x_0$ image - usually a tensor of zeros - is generated and a set of interpolated images are computed according to the formula $x_i=x_0+\alpha (x-x_0)$ where $x$ is the actual image that we are trying to explain, $\alpha$ is a set linearly spaced of coefficients in $[0,1]$.
All those images are passed to the network and the gradients along the path to the chosen class are collected and integrated.
Then, we smoothed the images with a 3D Gaussian kernel with $\sigma=4$ to reduce noise in the procedure and keep the values above the $95$ percentile to get a mask for the most important regions.
We repeated the procedure $10$ times sampling each time a slightly different neural network and then averaging the attributions masks.

\section{Results}

As a first experiment, we tested the standard neural network.
In this case, on the $100\%$ of the test set we obtained an accuracy of $0.86$, F1-score of $0.87$, precision $0.86$ and recall $0.86$ with an AUC of $0.938$.

\begin{figure}[h]
    \centering
    \includegraphics[width=0.45\textwidth]{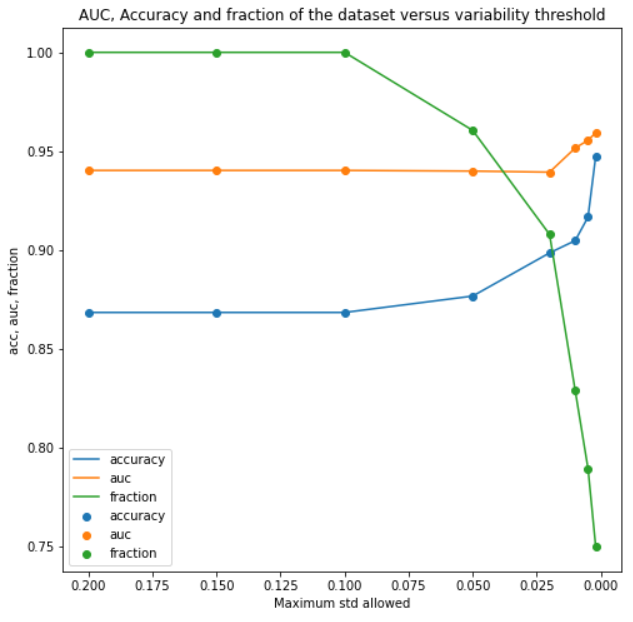}
    \includegraphics[width=0.45\textwidth]{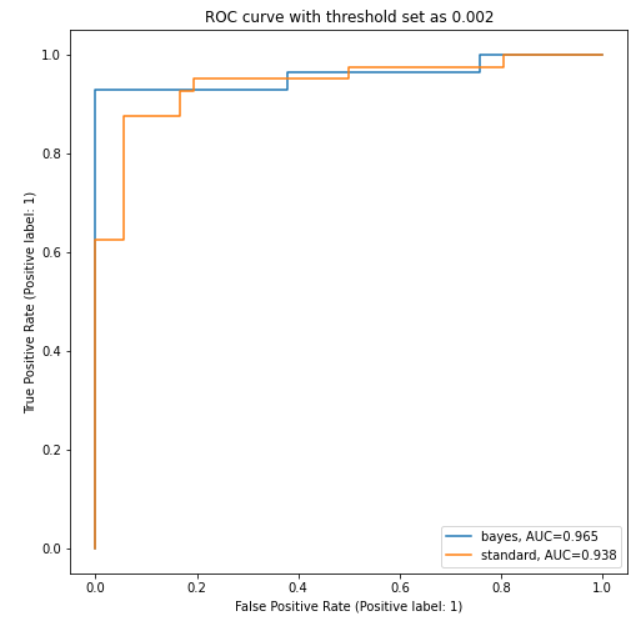}
    \caption{Results: In the left figure accuracy (blue), AUC (orange), and the fraction of the test dataset (green) are shown as functions of the threshold. Right figure: AUC for the standard model and a selected bayesian model with a threshold of 0.002.  }
    \label{fig:results}
\end{figure}

\begin{figure}[h!]
    \centering
    \includegraphics[width=0.9\linewidth]{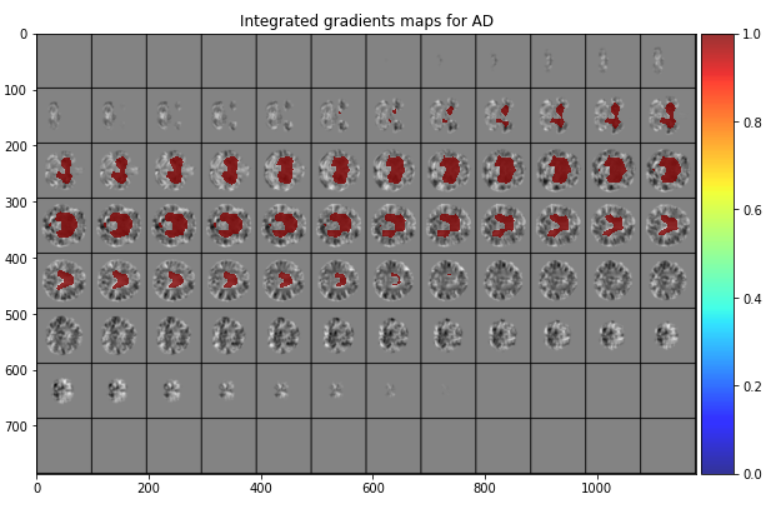}

    \caption{Results: Integrated Gradients. This is an interoperability method to look at the most influent areas for prediction. In this case, the model focus is on the ventriculi of the brain, an area which is involved in neurodegeneration.}
    \label{fig:integrad}
\end{figure}

Successively, our approach was tested as a function of t (i.e. the maximum standard deviation accepted for the class with the highest probability, see above). WE compured area under the receiver operating characteristic curve, and the fraction of the retained test dataset for each threshold. In Fig \ref{fig:results}, the results are reported as a function of the threshold value. We can clearly see two opposite trends, where reducing the threshold the accuracy and AUC increase while the fraction of the remaining test dataset naturally decreases since the model is rejecting the predictions whose associated uncertainty exceeds the threshold. In our use case, we observed the best results with a threshold of $0.002$, which retains $75\%$ of the dataset and reaches an accuracy of $0.95$ and AUC of $0.96$. Figure \ref{fig:integrad} shows the final explainability masks generated by averaging integrated gradients for randomly chosen AD and healthy cases in the test set. It appears that the model focuses on different areas of the lower brain and, in particular, the ventricular spaces, whose deformation is known to be correlated with AD  \cite{ventriculi,enlargment}.

\begin{table}[htbp]
\label{tab:results}
\centering
\large
\begin{tabular}{@{\hskip 10pt}l@{\hskip 12pt}l@{\hskip 10pt}l@{\hskip 10pt}l}
\multicolumn{1}{l|}{\textbf{Threshold}} & \textbf{Accuracy} & \textbf{AUC} & \textbf{fraction} \\ \hline
\multicolumn{1}{l|}{0.002}              & \textbf{0.947}             & \textbf{0.959}        & 0.750             \\
\multicolumn{1}{l|}{0.005}              & 0.916             & 0.955        & 0.789             \\
\multicolumn{1}{l|}{0.01}               & 0.904             & 0.951        & 0.829             \\
\multicolumn{1}{l|}{0.02}               & 0.898             & 0.939        & 0.907             \\
\multicolumn{1}{l|}{0.05}               & 0.876             & 0.939        & 0.960             \\
\multicolumn{1}{l|}{0.10}               & 0.868             & 0.940        & 1.0               \\
\multicolumn{1}{l|}{0.15}               & 0.868             & 0.940,       & 1.0               \\
\multicolumn{1}{l|}{0.2}                & 0.868             & 0.940        & 1.0               \\
                                        &                   &              &                  
\end{tabular}
\caption{Results: Accuracy, AUC, and the fraction of the test dataset with uncertainty behind the threshold.}
\end{table}

\section{Discussion}

The willingness to trust predictions by automatic devices, especially those based on black-box algorithms like neural networks, is critical in a vast number of application domains, such as e.g. the medical field. Ethics and responsibility pose an upper bound on the contribution of such techniques in medical diagnosis, screening, and triaging. We believe that being able to estimate the uncertainty of a prediction, along with tools that can modulate the behavior of the network to the degree of confidence that the user is informed about (and comfortable with) can represent a crucial step in this direction. Such features can improve the translation from research to clinical predictive models. Rather than completely replacing humans in evaluation, AI can support extremely useful recommendation systems and powerful tools to reduce workload in an efficient way for e.g. medical professionals. We proposed a method that turns a classical neural network into a Bayesian neural network, hence endowing the model with the ability to estimate the uncertainty associated with predictions. We also incorporate a rejection method based on a threshold based on thresholding the estimated uncertainty, which has resulted in a global performance increase (which amounts to reducing probably misclassified cases as they are associated with higher uncertainty). Additionally, by exclusion, this system can select cases to be recommended for expert human evaluation when the uncertainty is above the threshold.

\section{Conclusion}
We built a Bayesian-based neural network method able to estimate variability in predictions by simulating sampling from an infinite neural network ensemble. We used the estimated variability combined with a rejection method to retain only the fraction of the dataset that the model is able to classify with an under-threshold uncertainty, and showed that this procedure can improve the accuracy from $0.86$ to $0.95$ (while retaining $75\%$ of the test) when discriminating for AD from healthy cases based on brain morphometry only. Using integrated gradients, we also found that our model focuses on brain areas that are consistent with the clinical presentation of AD, in addition to highlighting previously unexplored areas in the lower part of the brain.

\section*{Acknowledgements}

{\footnotesize
Part of this work is supported by the EXPERIENCE project (European Union’s Horizon 2020 research and innovation program under grant agreement No. 101017727)

Matteo Ferrante is a PhD student enrolled in the National PhD in Artificial Intelligence, XXXVII cycle, course on Health and life sciences, organized by Università Campus Bio-Medico di Roma.

}

%
%
%
\bibliographystyle{splncs04}
\bibliography{references.bib}

\end{document}